# Cavitation near an oscillating piezoelectric plate in water

M. N. Shneider[1], M. Pekker[2]

[1]*Department of Mechanical and Aerospace Engineering, Princeton University, Princeton, NJ 08544*

[2]*Retired, Independent researcher, Princeton, NJ 08540*

## Abstract

Gas and vapor bubbles at a wall can reduce the skin friction of a liquid flow. We estimate whether an oscillating piezoelectric plate mounted flush with the wall can create such a near-wall bubble layer by acoustic cavitation. Once the cavitation threshold is reached, the liquid can no longer follow the receding plate, and the excess plate displacement is accommodated by cavitation bubbles forming at the wall; this kinematic argument gives the bubble volume generated per stroke and its energy cost. Homogeneous cavitation in pure water (about −30 MPa) requires pressure amplitudes far beyond those of flat piezoelectric plates and produces sub-micron bubbles that collapse within nanoseconds. Heterogeneous cavitation on wall and gas nuclei (−0.1 to −1 MPa) is within reach of piezoceramics and concentrates the void on fewer, larger wall-attached bubbles, which survive between strokes if they exceed a few micrometres at 5 MHz. The acoustic power needed to sustain a bubble layer in a turbulent boundary layer at 10 m/s is at least about 20 kW/m² for heterogeneous and 0.4 MW/m² for homogeneous cavitation, if the bubbles are stabilized by gas between bursts of strokes; continuous driving raises it by an order of magnitude or more.

\

## 1. Introduction

Injecting gas bubbles into a turbulent boundary layer can reduce skin friction substantially, provided the void fraction near the wall is large enough and the bubbles reside in the buffer layer [1,2]. A related effect is the slip produced by gas pockets trapped on structured or superhydrophobic surfaces ("bubble mattresses") [3,4]; slip can turn into extra drag when bubbles protrude too far [3]. Drag reduction in two-phase near-wall flows is also discussed in [5].

Here we ask whether such a layer can be produced without gas injection, by cavitation driven from the wall itself. A piezoelectric plate mounted flush with the wall radiates ultrasound into the liquid (Fig. 1). During the receding phase of the plate, the liquid at the wall is put under tension. If the tension exceeds the cavitation threshold, bubbles form in or at the wall.

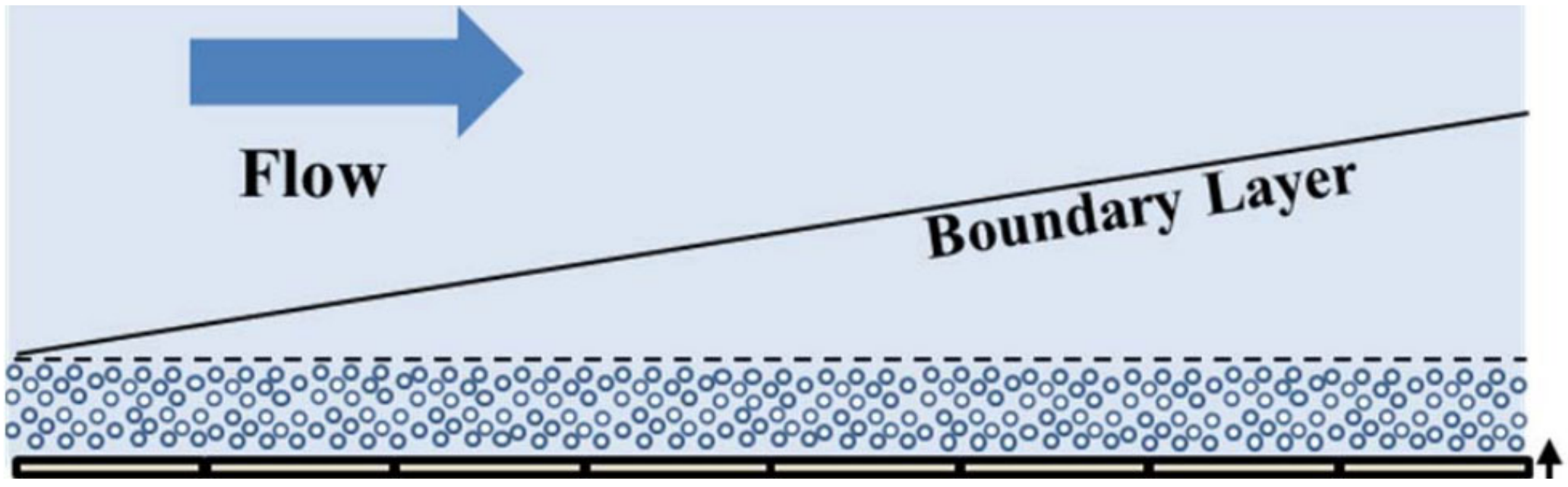

**Fig. 1.** Scheme of cavitation generation by an array of oscillating piezoelectric plates in the wall under a boundary layer. The vertical arrow shows the plate displacement; the dotted region marks where bubbles are generated, which coincides with the region where the sound wave is absorbed by the bubbles.

# 2. Acoustic field of the plate

**Linear piston model.** Consider normal (piston-like) vibrations of a flat plate at the wall. In the linear approximation without dissipation [6]:

$$\frac{\partial \rho'}{\partial t} + \rho_0 \frac{\partial v'}{\partial y} = 0, \qquad \frac{\partial v'}{\partial t} = -\frac{1}{\rho_0}\frac{\partial P'}{\partial y}, \qquad P' \approx c_s^2 \rho' \tag{1}$$

Here $\rho'$, $P'$, $v'$ are the perturbations of density, pressure and wall-normal velocity, $y$ is the distance from the wall, $\rho_0 = 10^3$ kg/m³ and $c_s \approx 1500$ m/s. The linear equation of state is adequate for the density at $|P'| \lesssim 10^2$ MPa [7,8]. For plate displacement $a(t) = a_0 \sin(\omega t)$, measured into the liquid, the normal velocity of the liquid at the wall equals that of the plate. The pressure and velocity then vary as $\cos(\omega t - ky)$, $k = \omega/c_s$, with wall amplitudes

$$v_0 = \omega a_0, \qquad P_0 = \rho_0 c_s \omega a_0 = \rho_0 c_s v_0 \tag{2}$$

For example, $P_0 = 50$ MPa at $f = 5$ MHz requires $a_0 \approx 1.1$ µm and $v_0 \approx 33$ m/s. The plate is treated as an infinite piston: its size must greatly exceed the acoustic wavelength $\lambda_s = c_s/f$ (0.3 mm at 5 MHz, 7.5 mm at 200 kHz); otherwise the near-field structure of the radiation must be resolved.

**Attenuation.** The classical absorption coefficient of a plane wave is

$$\alpha = \frac{\omega^2}{2\rho_0 c_s^3}\left(\frac{4}{3}\eta + \zeta\right) \tag{3}$$

with shear viscosity $\eta \approx 10^{-3}$ Pa·s and bulk viscosity $\zeta \approx 2.8\eta$ for water [9]. This gives $\alpha/f^2 \approx 2.4 \cdot 10^{-14}$ s²/m, i.e. an absorption length $1/\alpha \approx 40$ m at 1 MHz, 1.7 m at 5 MHz and 4 mm at 100 MHz. Viscous absorption therefore cannot confine the cavitation to a thin near-wall layer. Confinement must come from the cavitation itself (Sections 3–4).

**Nonlinear distortion.** At the amplitudes of interest the wave steepens quickly. The shock-formation distance of a plane harmonic wave is [10]

$$y_{sh} = \frac{\rho_0 c_s^3}{\beta \omega P_0}, \qquad \beta = 1 + \frac{B}{2A} \approx 3.5 \tag{4}$$

Here $A = \rho_0 c_s^2$ and $B = \rho_0^2 (\partial^2 p/\partial \rho^2)_s$ are the coefficients of the adiabatic equation of state, $p - p_\infty = A(\rho'/\rho_0) + (B/2)(\rho'/\rho_0)^2 + \cdots$, and $B/A \approx 5$ for water [10]. For $P_0 = 50$ MPa at 5 MHz, $y_{sh} \approx 0.6$ mm $\approx 2\lambda_s$, so linear acoustics holds only close to the wall. Cavitation clips only the tensile phase at $P_{cr}$ (Section 4). Under harmonic driving the compressive phase keeps the full amplitude $P_0$ and can form a weak shock ($P_0/\rho_0 c_s^2 \approx 0.02$ even at 50 MPa) at $y \approx y_{sh}$, provided $y_{sh}$ is smaller than the absorption length $1/\alpha$ and the near-field length $l^2/\lambda_s$ of a plate of streamwise length $l$, whereas for a fast-receding, slow-return stroke only the weak compressive tail steepens. For $P_0 \approx 1.5$ MPa, $y_{sh} \approx 2$ cm at 5 MHz.

**Source limits.** The vibration velocity of hard PZT ceramics is typically limited to about 1 m/s by heating and depolarization [11]. By Eq. (2) this corresponds to $P_0 \approx 1.5$ MPa. Pressure amplitudes of tens of MPa are produced in practice only at the focus of focused transducers or by shock-wave sources, not by flat, wall-mounted plates. This limit is decisive for the choice of operating regime (Section 3).

# 3. Cavitation threshold

Cavitation can start in two ways. In homogeneous nucleation, vapor nanopores are created in the pure liquid itself by thermal fluctuations. A large tension is needed for a fluctuation to produce a pore of critical size $R_{cr}$ of a few nanometres; pores appear throughout the stretched volume, and their number is set by the nucleation rate. In heterogeneous nucleation, pre-existing nuclei grow: gas microbubbles in the liquid, gas trapped in crevices of the wall, hydrophobic spots or suspended particles. No thermally activated barrier has to be overcome; the liquid only has to be stretched below the Blake threshold of the nucleus, and the number of bubbles is set by the number of nuclei. Cavitation always starts at the lower of the two thresholds, which in technical water near a real wall is the heterogeneous one.

## 3.1. Homogeneous nucleation

In pure, degassed water the measured cavitation pressure is $P_{cr} \approx -30$ MPa [12]. The nucleation rate per unit volume in liquid under tension is [13–15]

$$J = \frac{dn_b}{dt} \simeq \frac{1}{V_b}\frac{k_B T}{2\pi\hbar}\exp\left(-\frac{W_b}{k_B T}\right), \qquad V_b = \frac{4}{3}\pi R_b^3 \tag{5}$$

The pre-exponential factor differs between authors [14,15] but enters the results only logarithmically. With the radius-dependent surface tension, taken in Lorentzian form [13] for convenience,

$$\sigma = \frac{\sigma_0 R_b^2}{R_b^2 + \delta_b^2} \tag{6}$$

where $\sigma_0 = 0.072$ N/m [16] and $\delta_b$ is a Tolman-like length [17], the work of forming a bubble is [13]

$$W_b = -\frac{4\pi}{3}|P_-|R_b^3 + 4\pi R_b^2\sigma_0\left[1 - \frac{\delta_b^2}{R_b^2}\ln\left(1 + \frac{R_b^2}{\delta_b^2}\right)\right] \tag{7}$$

Its maximum at fixed tension $|P_-|$ defines the critical radius

$$R_{cr} = \frac{\sigma_0}{|P_-|} + \sqrt{\left(\frac{\sigma_0}{|P_-|}\right)^2 - \delta_b^2} \tag{8}$$

The experimental value $P_{cr} = -30$ MPa is reproduced with $\delta_b \approx 2.26$ nm, which gives $R_{cr} \approx 3.2$ nm. At this tension the barrier is $W_b(R_{cr}) \approx 15\, k_B T$. With $\delta_b = 0$ the same barrier would be reached at $P_{cr} \approx -320$ MPa; the often quoted value of about −1000 MPa corresponds to the criterion $W_b(R_{cr}) \approx k_B T$ [13].

**Saturation of the nanopore density.** Nucleation relieves the tension that drives it [13,18]. Pores of volume $V_{cr} = 4\pi R_{cr}^3/3$ at number density $n_b$ reduce the tension by $\delta P \approx \rho_0 c_s^2 V_{cr} n_b$. A hard upper bound on $n_b$ follows from $\delta P \le |P_{cr}|$:

$$n_b \le \frac{|P_{cr}|}{\rho_0 c_s^2 V_{cr}} \approx 10^5\ \mu\text{m}^{-3} \tag{9}$$

Because the rate falls exponentially as the tension relaxes, the actual density is much lower [13]:

$$n_{b,sat} = \frac{k_B T}{\rho_0 c_s^2 V_{cr}^2} \ln\left(1 + \frac{\rho_0 c_s^2 V_{cr}^2}{k_B T} J(|P_{cr}|)\, \delta t_-\right) \qquad (10)$$

Eq. (10) would reach the bound (9) only if the logarithm were about $|P_{cr}|V_{cr}/k_B T \approx 10^3$, which is unattainable. For the parameters above, $n_{b,sat} \approx 10^3$ μm⁻³ after tens of nanoseconds (Fig. 2). The interval during which the wall pressure is below $P_{cr}$ is

$$\delta t_- = \frac{2}{\omega} \arccos \frac{|P_{cr}|}{P_0} \qquad (11)$$

which at 5 MHz equals 59, 90 and 95 ns for $P_0 = 50$, 200 and 370 MPa.

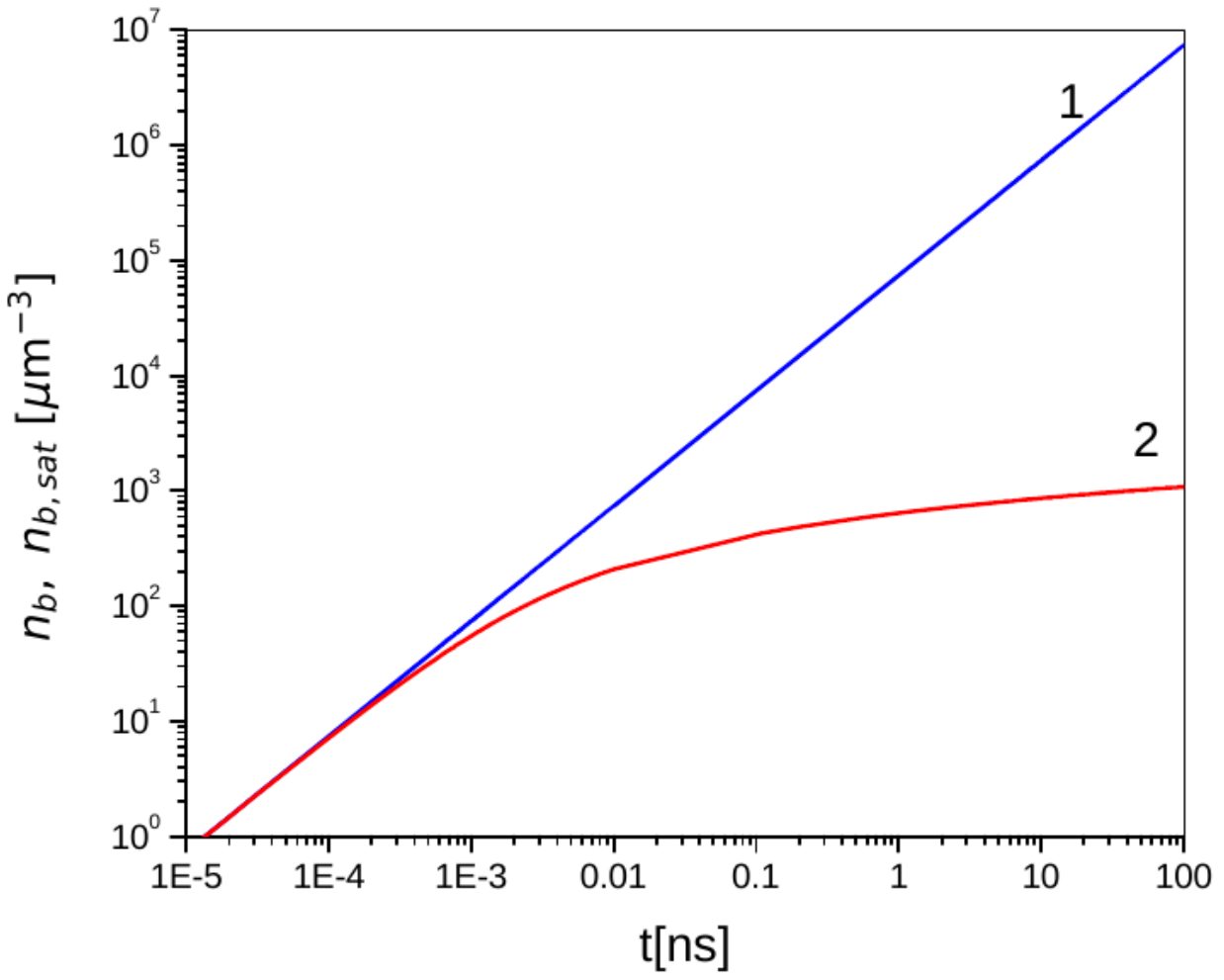

**Fig. 2.** Nanopore number density versus time at constant $p = P_{cr} = -30$ MPa, $\delta_b = 2.26$ nm. Curve 1: without tension relief; curve 2: with relief, Eq. (10).

## 3.2. Heterogeneous nucleation and practical thresholds

Technical water and real walls contain cavitation nuclei: gas microbubbles, particles, and gas trapped in surface crevices and on hydrophobic spots [19,20]. A gas nucleus of equilibrium radius $R_0$ grows without bound when the liquid pressure drops below the Blake threshold [19]

$$P_B = p_v - \frac{4\sigma_0}{3}\left[\frac{2\sigma_0}{3R_0^3\,(p_\infty + 2\sigma_0/R_0 - p_v)}\right]^{1/2} \approx p_v - 0.77\,\frac{\sigma_0}{R_0} \qquad (12)$$

where $p_v \approx 2.3$ kPa is the vapor pressure and $p_\infty \approx 0.1$ MPa the ambient pressure. The approximate form holds for $2\sigma_0/R_0 \gg p_\infty$. Eq. (12) gives $P_B \approx -5.5$, $-0.53$ and $-0.04$ MPa for $R_0 = 10$ nm, 0.1 μm and 1 μm.

Thus, near a real wall, cavitation is expected at $P_{cr} \approx -0.1$ to $-1$ MPa rather than −30 MPa. This range is compatible with the source limit $P_0 \approx 1.5$ MPa of Section 2. In this regime the bubble density is set by the density of nuclei, not by Eq. (10). For wall nuclei we characterize it by the number of active sites per unit wall area, $N_s$.

# 4. Kinematic limit: void opening at the receding wall

In a plane wave the pressure and velocity are related by $P' = \rho_0 c_s v'$. A liquid that can sustain a tension of at most $|P_{cr}|$ can therefore follow a receding wall with a velocity of at most

$$v_* = \frac{|P_{cr}|}{\rho_0 c_s} \qquad (13)$$

which is 20 m/s for $|P_{cr}| = 30$ MPa and 0.67 m/s for $|P_{cr}| = 1$ MPa. When the plate recedes faster, $v_p(t) > v_*$ (here and below $v_p > 0$ is the velocity of the plate away from the liquid), the transmitted wave is clipped at $P_{cr}$, and the excess displacement has to be filled by cavities. This is the one-dimensional mechanism of column separation known from water hammer [21]. The void volume opened per unit wall area per stroke is

$$h_v = \int \left[v_p(t) - v_*\right]_+ dt \qquad (14)$$

For harmonic motion, $v_p = v_0 \cos\omega t$ during the receding half-period ($t$ counted from its middle), and with $q = v_*/v_0 = |P_{cr}|/P_0$:

$$h_v = \frac{2}{\omega}\left[v_0\sqrt{1-q^2} - v_* \arccos q\right] \qquad (15)$$

The tension is maximal at the wall, where the wave originates, and the nuclei of Section 3.2 sit on the wall. Hence the void opens at, or within a fraction of a micrometre of, the wall. This is the physical origin of the localization sketched in Fig. 1. It does not rely on viscous absorption, which is negligible (Section 2).

**Static pressure.** $P_0$ is the acoustic pressure amplitude, added to the static pressure $p_\infty$, whereas $P_{cr}$, like the Blake threshold (12), is an absolute pressure. The absolute pressure at the wall is $p = p_\infty - \rho_0 c_s |v_p|$, so cavitation starts when $P_0 > p_\infty + |P_{cr}|$, and strictly $v_* = (p_\infty + |P_{cr}|)/(\rho_0 c_s)$ and $q = (p_\infty + |P_{cr}|)/P_0$. For homogeneous nucleation the difference (30.1 instead of 30 MPa) is negligible. In Sections 4–6, $|P_{cr}|$ therefore stands for the threshold acoustic amplitude $p_\infty + |P_{cr}|$. For the heterogeneous cases of Tables 1–3, the values 1 and 0.5 MPa correspond to absolute thresholds of −0.9 and −0.4 MPa.

**Energy partition.** The work done by the plate per unit area per stroke, with the wall pressure clipped at $|P_{cr}|$, is

$$E_1 = \rho_0 c_s \int v_p \min(v_p, v_*)\, dt = \rho_0 c_s \int \min(v_p, v_*)^2\, dt + |P_{cr}|\, h_v \qquad (16)$$

The first term is radiated into the bulk as a clipped wave; the second is spent on opening void. The fraction spent on cavitation is

$$\varphi = \frac{|P_{cr}|\, h_v}{E_1} \qquad (17)$$

Eq. (16) replaces the energy balance of the first version, in which only the energy delivered during the bubble-growth time was counted.

Representative cases are listed in Table 1 (harmonic receding phase; computed from Eqs. (2), (11), (15)–(17)).

**Table 1.** Void opened per stroke and energy partition. Cases A, B: homogeneous nucleation; C–E: heterogeneous nucleation. $E_1$ includes only the receding half-period, i.e. a slow return stroke is assumed (see Section 6).

| Case | $\|P_{cr}\|$, MPa | $P_0$, MPa | $f$, MHz | $a_0$, µm | $v_*$, m/s | $h_v$, µm | $\delta t_-$, ns | $E_1$, J/m² | $\varphi$ |
|---|---|---|---|---|---|---|---|---|---|
| A | 30 | 50 | 5 | 1.06 | 20 | 0.52 | 59 | 60 | 0.26 |
| B | 30 | 200 | 5 | 4.2 | 20 | 6.6 | 90 | 254 | 0.78 |
| C | 1 | 1.5 | 5 | 0.032 | 0.67 | 0.012 | 54 | 0.059 | 0.20 |
| D | 1 | 1.5 | 0.2 | 0.80 | 0.67 | 0.29 | 1340 | 1.46 | 0.20 |
| E | 0.5 | 1.5 | 0.2 | 0.80 | 0.33 | 0.85 | 1960 | 0.78 | 0.54 |

At a fixed vibration velocity, $h_v$ scales as $1/f$. Lower frequencies therefore open more void per stroke. The price is a larger plate (size $\gg \lambda_s$) and, as shown next, a stricter condition for bubble survival.

**Wettability.** The kinematic argument fixes only the void volume $h_v$, not where the void appears. The liquid does not detach from the wall as a continuous gap. On a well-wetted wall the work of adhesion $\sigma_0(1+\cos\theta)$, where $\theta$ is the contact angle of water, is close to the work of cohesion $2\sigma_0$, and the void opens as bubbles in the liquid within a short distance of the wall. On a hydrophobic wall bubbles form on the wall itself. The much lower heterogeneous thresholds of Section 3.2 are due to gas trapped in crevices and on hydrophobic spots, which hydrophobicity stabilizes. Wettability and roughness of the plate surface therefore control the number of active sites $N_s$ and the bubble shape. The hemispherical bubbles assumed in Eq. (23) correspond to $\theta \approx 90°$, while hydrophobic walls give flattened, attached bubbles close to the bubble mattresses of [3].

# 5. Bubble growth and survival between strokes

## 5.1. Growth during the tensile phase

For $R_b \ll \lambda_s$ and bubbles far apart compared with their size, each bubble obeys the Rayleigh equation in the local liquid pressure $p(t)$ [13,22]:

$$\frac{d}{dt}\left[R_b^3\left(\frac{dR_b}{dt}\right)^2\right] = \frac{2}{\rho_0}\left[-p(t) - \frac{2\sigma_0}{R_b\,(1+\delta_b^2/R_b^2)}\right]R_b^2\,\frac{dR_b}{dt} \qquad (18)$$

Once $R_b \gg R_{cr}$, surface tension is negligible under constant tension $p \approx P_{cr}$, and the radius grows linearly:

$$R_b \approx v_b t, \qquad v_b = \left(\frac{2|P_{cr}|}{3\rho_0}\right)^{1/2} \qquad (19)$$

This gives $v_b \approx 140$ m/s for $|P_{cr}| = 30$ MPa and 26 m/s for 1 MPa.

**Homogeneous case.** With $n_{b,sat} \approx 10^3$ µm⁻³, the bubble void fraction $\xi = (4\pi/3)\,n_{b,sat}(v_b t)^3$ grows until the strain of the wave is taken up. In the first version this was expressed through the energy balance $\xi|P_{cr}| = P_0^2/(2\rho_0 c_s^2)$, i.e.

$$\xi = \frac{P_0^2}{2\rho_0 c_s^2|P_{cr}|} = 7.4\cdot 10^{-6}\,P_0^2\ [\text{MPa}] \qquad (20)$$

The value $7.4\cdot 10^{-6}$ is consistent with $\xi = 1$ at $P_0 \approx 367$ MPa. The tension is relieved within $t_{s,d} \approx$ 0.1–0.5 ns, when the bubbles have reached only $R_b \approx v_b t_{s,d} \approx 16$ nm (at $P_0 = 50$ MPa).

The wave energy is therefore absorbed within $c_s t_{s,d} \approx$ 0.15–0.75 μm of the wall, in agreement with the kinematic picture of Section 4.

## 5.2. Collapse after the tensile phase

After the tension is relieved, a cavity of radius $R_m$ is compressed by $\Delta p = p_\infty - p_v + 2\sigma_0/R_m$. For an incompressible liquid the radial velocity is $w(r) = R^2\dot{R}/r^2$, and the kinetic energy of the liquid around the cavity is $2\pi\rho_0 R^3\dot{R}^2$. For a cavity starting from rest at $R = R_m$, equating this energy to the work of the pressure difference gives the first integral of the Rayleigh equation [22] and the collapse time:

$$\dot{R}^2 = \frac{2\Delta p}{3\rho_0}\left(\frac{R_m^3}{R^3} - 1\right), \qquad t_c = \int_0^{R_m} \frac{dR}{|\dot{R}|} = R_m\left(\frac{3\rho_0}{2\Delta p}\right)^{1/2} \int_0^1 \frac{z^{3/2}\,dz}{\sqrt{1-z^3}}$$

The last integral equals $\Gamma(5/6)\,\Gamma(1/2)/[3\,\Gamma(4/3)] \approx 0.747$, where $\Gamma$ is the gamma function, which yields the Rayleigh collapse time of an empty cavity [19,22]:

$$t_c \approx 0.915\,R_m\left(\frac{\rho_0}{\Delta p}\right)^{1/2} \qquad (21)$$

With the size-dependent $\sigma(R)$ of Eq. (6), the exact collapse time differs from Eq. (21) by less than 5%; gas content inside the bubble would lengthen it. For the 16 nm homogeneous bubbles present when the tension is first relieved, $2\sigma(R_m)/R_m \approx 9$ MPa and $t_c \approx 1.5 \cdot 10^{-10}$ s. The plate, however, keeps receding during the whole interval $\delta t_-$, and the bubbles adjacent to the wall keep growing. Sharing the void $h_v \approx 0.52$ μm (Table 1, case A) among about $n_{b,sat}^{2/3} \approx 10^{14}$ m$^{-2}$ pores of the first layer gives $R_m \approx 0.1$ μm at the end of the stroke and $t_c \approx 2 \cdot 10^{-9}$ s. Even these bubbles disappear about 100 times faster than the stroke period $T_r = 0.2$ μs and cannot accumulate.

The survival condition given in the first version, $R_b > 2\sigma_0/p_v$, is withdrawn. It omits the ambient pressure: a purely vapor-filled bubble with $p_v < p_\infty$ collapses whatever its size. The correct requirement is that a bubble does not collapse before the next stroke, which arrives after the repetition period $T_r \geq 1/f$. Neglecting the Laplace term, $\Delta p \approx p_\infty - p_v$, the condition $t_c(R_m) > T_r$ gives

$$R_m > R_* \approx \frac{T_r}{0.915}\left(\frac{p_\infty - p_v}{\rho_0}\right)^{1/2} \qquad (22)$$

Eq. (22) is accurate for $R_* \gg 2\sigma_0/p_\infty \approx 1.4$ μm. With the Laplace term retained, $R_*$ is obtained by substituting $\Delta p = p_\infty - p_v + 2\sigma_0/R_*$ into $t_c(R_*) = T_r$. This gives the cubic $R_*^3 - A(p_\infty - p_v)R_* - 2A\sigma_0 = 0$, where $A = (T_r/0.915)^2/\rho_0$, which has a single positive root. For $T_r = 1/f$ this gives $R_* \approx 2.7$ μm at 5 MHz (Eq. (22) gives 2.2 μm), 54 μm at 200 kHz and 0.22 mm at 50 kHz.

Condition (22) is sufficient but not necessary for accumulation from stroke to stroke. During collapse the vapor condenses at the bubble wall. In cold water the released latent heat does not slow the inertial collapse appreciably, and it cannot produce a new vapor bubble. The work of the ambient pressure, about $p_\infty V_m$ with $V_m = 4\pi R_m^3/3$, is more than an order of magnitude smaller than the latent heat $\rho_v L \approx 1.4 \cdot 10^6$ J/m³ needed to fill the same volume with saturated vapor at 100 °C ($\rho_v \approx 0.6$ kg/m³ is the vapor density and $L \approx 2.26 \cdot 10^6$ J/kg the latent heat of vaporization). The heated layer also cools within $R_m^2/\chi$ ($\chi \approx 1.4 \cdot 10^{-7}$ m²/s is the thermal diffusivity of water), about 2 ns for $R_m = 16$ nm. Strong heating occurs only in the non-condensable gas and in the vapor trapped at the final stage of collapse; it makes the bubble rebound and perform several damped oscillations [19,20]. If dissolved gas has diffused into the bubble, a residual gas microbubble survives these oscillations. Such residual nuclei lower the cavitation threshold of the next stroke ("cavitation memory" [20]). Accumulation can therefore proceed through nuclei even when the bubbles themselves collapse between strokes.

A bubble that retains gas oscillates about its equilibrium radius $R_{eq}$ with the natural (Minnaert) frequency corrected for surface tension [19,20]:

$$\omega_0^2 = \frac{1}{\rho_0 R_{eq}^2}\left[3\kappa\left(p_\infty + \frac{2\sigma_0}{R_{eq}} - p_v\right) - \frac{2\sigma_0}{R_{eq}}\right]$$

where $\kappa$ is the polytropic exponent. For $R_{eq} \lesssim 3$ µm the thermal Péclet number $\omega_0 R_{eq}^2/\chi_g$ ($\chi_g \approx 2 \cdot 10^{-5}$ m²/s is the thermal diffusivity of air) is of order unity or less, so the oscillations are nearly isothermal and $\kappa \approx 1$ [25]. This gives $f_0 \approx 3.8$ MHz for $R_{eq} = 1$ µm and 1.05 MHz for 3 µm, with quality factors of order 3–7 owing to viscous, thermal and radiation damping. Bubbles with $R_{eq} \approx 0.8$ µm are therefore resonant with a 5 MHz drive, which favors their growth by rectified diffusion [24]. The phase of their oscillation at the arrival of the next stroke matters and has to be resolved numerically.

## 5.3. Wall-attached bubbles from heterogeneous sites

If the void $h_v$ of one stroke is shared among $N_s$ active wall sites per unit area, each forms a hemispherical bubble of radius

$$R_1 = \left(\frac{3h_v}{2\pi N_s}\right)^{1/3} \qquad (23)$$

**Table 2.** Radius of wall-attached bubbles after one stroke, Eq. (23), compared with the survival radius, Eq. (22).

| Case (Table 1) | $h_v$, µm | $R_1$ at $N_s = 10^8$ m⁻², µm | $R_1$ at $10^9$ m⁻², µm | $R_1$ at $10^{10}$ m⁻², µm | $R_*$ ($T_r = 1/f$), µm |
|---|---|---|---|---|---|
| A, 5 MHz | 0.52 | 13.5 | 6.3 | 2.9 | 2.7 |
| C, 5 MHz | 0.012 | 3.9 | 1.8 | 0.84 | 2.7 |
| D, 0.2 MHz | 0.29 | 11 | 5.2 | 2.4 | 54 |
| E, 0.2 MHz | 0.85 | 16 | 7.4 | 3.4 | 54 |

Two conclusions follow. First, accumulation is possible only if the void concentrates on relatively few sites, which is why heterogeneous nucleation matters: homogeneous nucleation spreads it over about $10^{14}$ bubbles per square metre. Second, since $R_* \propto 1/f$ while $R_1 \propto f^{-1/3}$, the ratio $R_1/R_*$ grows as $f^{2/3}$. Higher frequencies favor survival, while lower frequencies open more void per stroke. At 5 MHz, case C satisfies the survival condition only for $N_s \lesssim 10^8$ m⁻², and case A for $N_s$ up to $10^{10}$ m⁻², though with little margin there. At 200 kHz single-stroke bubbles collapse between strokes, since even the shortest repetition period, $T_r = 1/f = 5$ µs, exceeds their collapse time.

Surviving bubbles grow from stroke to stroke as a ratchet: each new tensile phase enlarges existing cavities instead of nucleating new ones. Gas diffusion during one stroke is negligible (diffusion length $\sqrt{D \cdot 100\ \text{ns}} \approx 15$ nm, where $D \approx 2 \cdot 10^{-9}$ m²/s is the diffusion coefficient of air in water). Over many strokes, however, rectified diffusion [24] supplies gas that stabilizes the bubbles against dissolution [23]. A quantitative description of this stage requires numerical simulation.

The type of boundary layer does not change the cavitation threshold. In a turbulent layer at $U = 10$ m/s the wall-pressure fluctuations are of order $(2–3)\tau_w \approx 0.3–0.5$ kPa, where $\tau_w \approx 150$ Pa is the wall shear stress (Section 6), two to three orders of magnitude below the acoustic tension. It does affect the supply of nuclei and gas and the residence time of bubbles. Turbulent mixing brings free nuclei and dissolved gas to the wall, which favors rectified diffusion, but ejections intermittently remove bubbles from the near-wall region. A laminar layer keeps bubbles over the plate longer, but nuclei and gas reach the wall only by slow convection and diffusion. In both cases shear does not detach micrometre bubbles from the wall, since $\eta SR/\sigma_0 \lesssim 10^{-2}$, where $S \lesssim 1.6 \cdot 10^5$ s$^{-1}$ is the wall shear rate (Section 7).

# 6. Number of strokes and power

Let $H$ be the void volume per unit wall area that the layer must hold, e.g. $H = \xi_t \delta_l$ for a target void fraction $\xi_t$ in a layer of thickness $\delta_l$. Assuming the bubbles survive between strokes (Section 5), the number of strokes needed is

$$N = \frac{H}{h_v} \qquad (24)$$

The bubbles are carried away by the near-wall flow with the mean streamwise velocity $u_w$ at the outer edge of the bubble layer, $y = \delta_l$, and must be renewed every $l/u_w$, where $l$ is the streamwise plate length. With Eqs. (16), (17) and (24), the time-averaged acoustic power per unit plate area is

$$\Theta = E_1 N \frac{u_w}{l} = \frac{|P_{cr}|\, H}{\varphi} \frac{u_w}{l} \qquad (25)$$

Eq. (25) has a transparent meaning. The minimum work to open void $H$ against tension $|P_{cr}|$ is $|P_{cr}|H$; it is divided by the efficiency $\varphi$ and multiplied by the renewal rate. The power depends on frequency and amplitude only through $\varphi$, and it scales directly with the threshold.

**Example.** We consider a turbulent boundary layer at a free-stream velocity $U = 10$ m/s. On a smooth wall, transition occurs at $Re_x \approx 5 \cdot 10^5$, i.e. only about 5 cm from the leading edge, so over almost the whole length of a hull or channel wall the layer is turbulent; this is also the regime in which microbubble drag reduction is effective [1,2]. We take a skin-friction coefficient $c_f \approx 0.003$. This value follows from the turbulent flat-plate correlation $c_f \approx 0.0592\, Re_x^{-1/5}$ at $Re_x = Ux/\nu \approx 3 \cdot 10^6$ ($x$ is the distance from the leading edge, and $\nu = \eta/\rho_0 \approx 10^{-6}$ m²/s is the kinematic viscosity of water at 20 °C), i.e. about 0.3 m from the leading edge; over 0.1–1 m, $c_f$ varies between 0.0037 and 0.0024. The friction velocity $u_\tau = (\tau_w/\rho_0)^{1/2}$, where $\tau_w = c_f \rho_0 U^2/2 \approx 150$ Pa is the wall shear stress, is then $u_\tau = U(c_f/2)^{1/2} \approx 0.4$ m/s and the viscous length $\ell_\nu = \nu/u_\tau \approx 2.5$ μm. Distances and velocities in wall units are $y^+ = yu_\tau/\nu$ and $u^+ = u/u_\tau$; a different $c_f$ changes $u_\tau$, $y^+$ and $u_w$ roughly as $c_f^{1/2}$. A layer $\delta_l = 30$ μm ($y^+ \approx 12$, lower buffer layer [1]) with $\xi_t = 0.3$ gives $H = 9$ μm. Further, $u_w \approx 4$ m/s ($u^+ \approx 10$ from the law of the wall) and $l = 1$ cm, so $u_w/l = 400$ s$^{-1}$. The results are given in Table 3.

**Table 3.** Number of strokes and acoustic power for $H = 9$ μm and $u_w/l = 400$ s$^{-1}$ (cases as in Table 1).

| Case | $\lvert P_{cr}\rvert$, MPa | $f$, MHz | $N$, Eq. (24) | Stroke rate $Nu_w/l$, s$^{-1}$ | $\Theta$, Eq. (25), kW/m² |
|---|---|---|---|---|---|
| A | 30 | 5 | 17 | $7 \cdot 10^3$ | 410 |

| Case | $\lvert P_{cr}\rvert$, MPa | $f$, MHz | $N$, Eq. (24) | Stroke rate $Nu_w/l$, s$^{-1}$ | $\Theta$, Eq. (25), kW/m² |
|---|---|---|---|---|---|
| B | 30 | 5 | 1.4 | $5.5\cdot10^2$ | 140 |
| C | 1 | 5 | 760 | $3\cdot10^5$ | 18 |
| D | 1 | 0.2 | 31 | $1.2\cdot10^4$ | 18 |
| E | 0.5 | 0.2 | 11 | $4\cdot10^3$ | 3.3 |

The homogeneous regime needs 0.1–0.4 MW/m² of acoustic power. More importantly, it needs pressure amplitudes far above what flat piezoelectric plates deliver (Section 2). The heterogeneous regime needs about 20 kW/m² or less, with amplitudes within the source limit.

Two assumptions make these values lower bounds. First, $E_1$ includes only the receding half-period, i.e. a slow return stroke. For harmonic driving (a resonant plate, Section 7) the compressive half-period radiates an additional $\rho_0 c_s v_0^2\pi/(2\omega)$ per stroke, which raises $\Theta$ by a factor of 2.3–2.4 in cases A, C and D and by 3.4–6 in cases E and B. Second, Eq. (25) assumes that the strokes are delivered in bursts of $N$ strokes with $T_r \approx 1/f$ once per renewal time $l/u_w$, and that the bubbles formed in a burst then persist, stabilized by gas, until they are carried away. Vapor bubbles that need a stroke every $T_r$ to survive (Section 5.2) require continuous driving with power $E_1 f$: about 0.3 MW/m² in cases C and D and about 300 MW/m² in case A. Shielding reduces this upper estimate: once the layer has formed, the plate works mainly on the existing bubbles, and less energy is radiated into the bulk.

These figures are acoustic powers. The electrical power is larger by the inverse transducer efficiency. The first version reported $\Theta \approx 22$ kW/m² for the homogeneous case; the corrected coefficient of that formula is $4.5\cdot10^{10}$, not $4.5\cdot10^{9}$. That estimate counted only the energy delivered during the bubble-growth time (about 0.5 ns) and so omitted the energy radiated into the bulk, which Eq. (16) includes.

The estimate (25) is uncertain in both directions. Near-wall bubbles move slower than $u_w$ if attached, which lowers the required power. But bubbles that do not survive a stroke (Section 5) must be recreated, which raises it.

# 7. Discussion and conclusions

**Relevance to drag reduction.** The mechanism of Section 4 produces bubbles attached to the wall. The relevant reference systems are therefore bubble mattresses and gas-pocket slip [3,4] rather than freely convected microbubbles [1,2]. For mattresses, the sign of the effect depends on bubble protrusion: small protrusion gives slip, while large protrusion increases drag [3]. Whether cavitation bubbles grown stroke by stroke can be kept in the favorable range is an open question. Heating of the liquid by bubble collapse and by transducer losses lowers its viscosity [26]. For a short heated plate in a turbulent boundary layer, the Lévêque solution [27] gives a heat-transfer coefficient $h_T \approx 0.81\,k_T\,(S/\chi l)^{1/3} \approx 2\cdot10^4$ W/(m²·K). Here $S = u_\tau^2/\nu \approx 1.6\cdot10^5$ s$^{-1}$ is the wall shear rate, $k_T$ and $\chi$ are the thermal conductivity and thermal diffusivity of water, and the thermal layer, about 20 μm thick, is comparable to the bubble layer. It extends beyond the viscous sublayer (about 12 μm) into the buffer layer, where turbulent mixing increases heat transfer, so this estimate of $h_T$ is conservative. With a wall heat flux of about $2\Theta$ (cavitation plus transducer losses), the wall overheating is about 1–2 K in the heterogeneous regime, which changes the viscosity by only a few per cent, but 20–40 K in the homogeneous regime and about 25–30 K for continuous driving in the heterogeneous regime ($\Theta \approx E_1 f$, Section 6).

**Erosion and noise.** Bubble collapse next to a wall produces liquid microjets and shock waves that erode surfaces [19,20]. The plate surface itself would be exposed to this. The lower amplitudes of the heterogeneous regime reduce, but do not remove, this risk. Broadband cavitation noise is an additional side effect for underwater applications.

**Pulse shape.** A fast receding stroke followed by a slow return (Fig. 3) avoids the compressive phase that collapses bubbles under harmonic driving. A resonant plate, however, moves almost harmonically. A sawtooth-like motion requires broadband, off-resonance driving or a combination of a fundamental and its harmonics, both of which reduce the attainable amplitude.

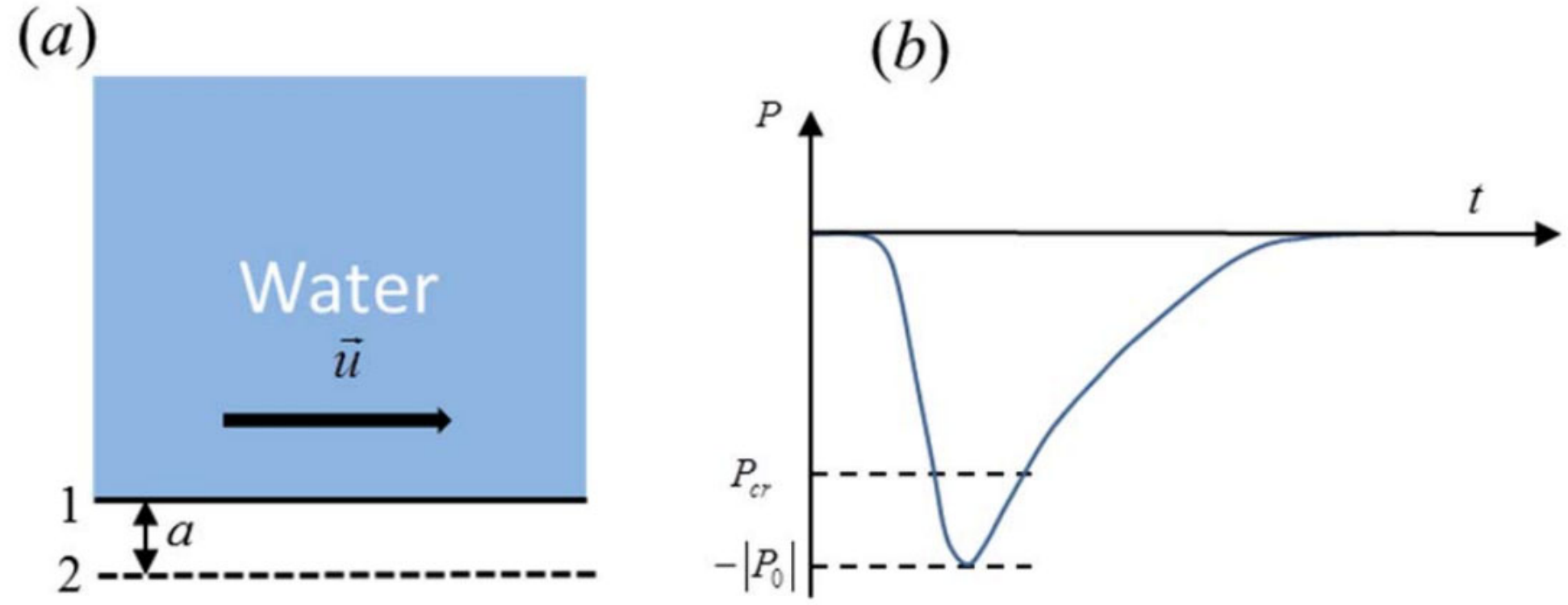

**Fig. 3.** (a) Oscillation of the plate surface: position 1 is the initial position, position 2 the maximum displacement. When the plate moves away from the liquid a tension appears; on the slow return no significant positive pressure arises. (b) Pressure in the liquid at the plate surface versus time.

**Limitations.** The analysis is one-dimensional and quasi-linear. It neglects the change of the mechanical properties of the bubbly liquid, bubble interaction and coalescence, gas diffusion beyond order-of-magnitude estimates, and the nonlinear propagation of the radiated wave. It provides estimates and conditions, not a quantitative description.

**Conclusions.**

1. A wall-mounted oscillating plate cavitates the liquid at the wall once the plate velocity exceeds $v_* = |P_{cr}|/(\rho_0 c_s)$. The void volume per stroke is given by Eq. (15), and a fraction $\varphi \approx 0.2$–$0.9$ of the stroke energy goes into cavitation.
2. Homogeneous cavitation ($P_{cr} \approx -30$ MPa) needs $P_0 > 30$ MPa (50 MPa in our example), far beyond flat piezoelectric plates, and produces sub-micron bubbles (about 0.1 µm at the end of the stroke) that collapse within a few nanoseconds.
3. Heterogeneous cavitation on wall nuclei ($P_{cr} \approx -0.1$ to $-1$ MPa) is reachable with $P_0 \approx 1.5$ MPa. It concentrates the void on $N_s$ sites per unit area.
4. Bubbles accumulate from stroke to stroke only if their radius exceeds $R_* \approx (T_r/0.915)(p_\infty/\rho_0)^{1/2}$, about 2 µm at 5 MHz, or about 3 µm when the Laplace pressure is included. This favors high frequency and a low density of active sites.
5. The power required to sustain a void layer $H$ is $\Theta \approx |P_{cr}| H u_w/(\varphi l)$. For a 30 µm layer with 30% void at 10 m/s this is about 20 kW/m² in the heterogeneous regime and about 0.4

MW/m² in the homogeneous one, provided the bubbles are stabilized by gas between bursts of strokes; continuous driving at $T_r \approx 1/f$ requires about $E_1 f$, i.e. about 0.3 MW/m² in the heterogeneous regime.

The key experimental questions are the density and threshold of active nucleation sites on the plate surface, the survival and growth of wall-attached bubbles under repeated strokes, and the resulting change of wall shear stress.

# Notation

| Symbol | Meaning |
|---|---|
| $x, y$ | Streamwise distance from the leading edge; distance from the wall (wall-normal coordinate) |
| $U$ | Free-stream velocity |
| $u_w$ | Mean streamwise flow velocity at the outer edge of the bubble layer, $y = \delta_l$ |
| $u_\tau = (\tau_w/\rho_0)^{1/2}$, $\ell_\nu = \nu/u_\tau$ | Friction velocity ($\tau_w$ is the wall shear stress; $u_\tau \approx 0.4$ m/s in Section 6); viscous length ($\approx 2.5$ μm) |
| $y^+ = yu_\tau/\nu$, $u^+ = u/u_\tau$ | Wall units ($u$ is the mean streamwise velocity at distance $y$) |
| $c_f$, $Re_x = Ux/\nu$ | Skin-friction coefficient; local Reynolds number |
| $l$ | Streamwise length of the plate |
| $a(t)$, $a_0$ | Plate displacement and its amplitude |
| $v_p(t)$, $v_0 = \omega a_0$ | Velocity of the plate away from the liquid (receding velocity) and its amplitude |
| $\rho'$, $P'$, $v'$ | Acoustic perturbations of density, pressure and wall-normal velocity |
| $P_0 = \rho_0 c_s v_0$ | Acoustic pressure amplitude at the wall |
| $\rho_0$, $c_s$ | Density of water and sound speed in water |
| $f$, $\omega = 2\pi f$, $k = \omega/c_s$, $\lambda_s = c_s/f$ | Frequency, angular frequency, wave number and wavelength of the plate oscillations |
| $\alpha$ | Acoustic absorption coefficient |
| $A$, $B$, $\beta = 1 + B/2A$ | Coefficients of the adiabatic equation of state; nonlinearity coefficient |
| $y_{sh}$ | Shock-formation distance |
| $p$, $p_\infty$, $p_v$ | Local absolute pressure; static (ambient) pressure; vapor pressure |
| $P_{cr}$ | Absolute cavitation pressure; in Sections 4–6, $\lvert P_{cr}\rvert$ stands for the threshold acoustic amplitude $p_\infty + \lvert P_{cr}\rvert$ |
| $P_-$, $P_B$ | Local tension in Eqs. (7), (8); Blake threshold, Eq. (12) |
| $\Delta p$ | Pressure difference driving bubble collapse |
| $V_m$, $\rho_v$, $L$ | Cavity volume at the start of collapse; density of saturated vapor; latent heat of vaporization |

| Symbol | Meaning |
|---|---|
| $J$, $n_b$, $n_{b,sat}$ | Nucleation rate per unit volume; nanopore number density and its saturated value |
| $W_b$, $\sigma$, $\sigma_0$, $\delta_b$ | Work of bubble formation; size-dependent and planar surface tension; Tolman-like length |
| $R_b$, $R_{cr}$, $V_b$, $V_{cr}$ | Bubble radius; critical radius; bubble and critical pore volume |
| $R_0$, $R_m$, $R_{eq}$, $R_1$, $R_*$ | Radius of a gas nucleus; radius at the start of collapse; equilibrium radius of a gas bubble; radius of a wall-attached bubble after one stroke; survival radius |
| $v_b$, $w(r)$ | Bubble-wall growth velocity; radial liquid velocity around a bubble |
| $\xi$, $\xi_t$, $\delta_l$ | Void fraction; target void fraction; thickness of the bubble layer |
| $\delta t_-$, $t_{s,d}$, $t_c$, $T_r$ | Duration of the tensile phase; tension-relief time; collapse time; stroke repetition period |
| $f_0$, $\omega_0$, $\kappa$ | Natural frequency of a gas bubble; polytropic exponent |
| $N_s$ | Number of active nucleation sites per unit wall area |
| $\theta$ | Contact angle of water on the wall |
| $v_*$, $q = v_*/v_0$ | Maximum velocity with which the liquid can follow the wall; threshold-to-amplitude ratio |
| $h_v$ | Void volume opened per unit wall area per stroke (equivalent layer thickness) |
| $E_1$, $\varphi$ | Work of the plate per unit area per stroke; fraction of it spent on cavitation |
| $H$, $N$ | Void volume per unit wall area to be maintained; number of strokes |
| $\Theta$ | Time-averaged acoustic power per unit plate area |
| $\eta$, $\zeta$, $\nu$ | Shear viscosity ($\approx 10^{-3}$ Pa·s), bulk viscosity ($\approx 2.8\eta$) and kinematic viscosity $\nu = \eta/\rho_0$ ($\approx 10^{-6}$ m²/s) of water at 20 °C |
| $D$, $\chi$, $\chi_g$ | Diffusion coefficient of air in water; thermal diffusivity of water and of air |
| $S$, $h_T$, $k_T$ | Wall shear rate; heat-transfer coefficient; thermal conductivity of water |
| $k_B$, $T$, $\hbar$ | Boltzmann constant; temperature; reduced Planck constant |

# References

[1] S. L. Ceccio, Friction drag reduction of external flows with bubble and gas injection, Annu. Rev. Fluid Mech. **42**, 183 (2010).

[2] Y. Murai, Frictional drag reduction by bubble injection, Exp. Fluids **55**, 1773 (2014).

[3] E. Karatay, A. S. Haase, C. W. Visser, C. Sun, D. Lohse, P. A. Tsai, R. G. H. Lammertink, Control of slippage with tunable bubble mattresses, Proc. Natl. Acad. Sci. USA **110**, 8422 (2013).

[4] J. P. Rothstein, Slip on superhydrophobic surfaces, Annu. Rev. Fluid Mech. **42**, 89 (2010).

[5] E. Ya. Gatapova, V. S. Ajaev, O. A. Kabov, On drag reduction in a two-phase flow, JETP Lett. **101**, 160 (2015).

[6] L. D. Landau, E. M. Lifshitz, *Fluid Mechanics*, 2nd ed. (Pergamon Press, Oxford, 1987).

[7] P. G. Tait, Report on some of the physical properties of fresh water and sea water, Rep. Sci. Results Voyage H.M.S. Challenger, Phys. Chem. **2**, 1 (1888).

[8] Y.-H. Li, Equation of state of water and sea water, J. Geophys. Res. **72**, 2665 (1967).

[9] M. J. Holmes, N. G. Parker, M. J. W. Povey, Temperature dependence of bulk viscosity in water using acoustic spectroscopy, J. Phys.: Conf. Ser. **269**, 012011 (2011).

[10] M. F. Hamilton, D. T. Blackstock (eds.), *Nonlinear Acoustics* (Academic Press, San Diego, 1998).

[11] K. Uchino (ed.), *Advanced Piezoelectric Materials: Science and Technology* (Woodhead Publishing, Cambridge, 2010).

[12] E. Herbert, S. Balibar, F. Caupin, Cavitation pressure in water, Phys. Rev. E **74**, 041603 (2006).

[13] M. N. Shneider, M. Pekker, *Liquid Dielectrics in an Inhomogeneous Pulsed Electric Field. Dynamics, Cavitation and Related Phenomena*, 2nd edn (IOP Publishing, Bristol, 2020).

[14] Ya. B. Zel'dovich, Theory of formation of a new phase. Cavitation, Zh. Eksp. Teor. Fiz. **12**, 525 (1942).

[15] J. C. Fisher, The fracture of liquids, J. Appl. Phys. **19**, 1062 (1948).

[16] W. M. Haynes (ed.), *CRC Handbook of Chemistry and Physics*, 93rd ed. (CRC Press, Boca Raton, 2012).

[17] R. C. Tolman, The effect of droplet size on surface tension, J. Chem. Phys. **17**, 333 (1949).

[18] M. Pekker, M. N. Shneider, Initial stage of cavitation in liquids and its observation by Rayleigh scattering, Fluid Dyn. Res. **49**, 035503 (2017).

[19] C. E. Brennen, *Cavitation and Bubble Dynamics* (Oxford University Press, New York, 1995).

[20] T. G. Leighton, *The Acoustic Bubble* (Academic Press, London, 1994).

[21] A. Bergant, A. R. Simpson, A. S. Tijsseling, Water hammer with column separation: A historical review, J. Fluids Struct. **22**, 135 (2006).

[22] Lord Rayleigh, On the pressure developed in a liquid during the collapse of a spherical cavity, Philos. Mag. **34**, 94 (1917).

[23] P. S. Epstein, M. S. Plesset, On the stability of gas bubbles in liquid–gas solutions, J. Chem. Phys. **18**, 1505 (1950).

[24] L. A. Crum, Rectified diffusion, Ultrasonics **22**, 215 (1984).

[25] A. Prosperetti, L. A. Crum, K. W. Commander, Nonlinear bubble dynamics, J. Acoust. Soc. Am. **83**, 502 (1988).

[26] J. Kestin, M. Sokolov, W. A. Wakeham, Viscosity of liquid water in the range −8 °C to 150 °C, J. Phys. Chem. Ref. Data **7**, 941 (1978).

[27] A. Lévêque, Les lois de la transmission de chaleur par convection, Ann. Mines **13**, 201–299, 305–362, 381–415 (1928).